\documentclass[%
 reprint,
 amsmath,amssymb,
 aps,
]{revtex4-1}

\usepackage{graphicx}
\usepackage{dcolumn}
\usepackage{bm}
\usepackage{array}
\usepackage[all,dvips]{xy}
\usepackage{epsfig}

\usepackage{epstopdf}

\begin{document}

\preprint{APS/123-QED}

\title{The smallest chimera states}

\author{Yuri Maistrenko$^{1,2}$,  Serhiy Brezetsky$^1$, Patrycja Jaros$^1$, Roman Levchenko$^2$, Tomasz Kapitaniak$^1$}

\affiliation{$^{1}$Division of Dynamics, Technical University of Lodz, Stefanowskiego 1/15, 90-924 Lodz, Poland}

\affiliation{$^{2}$Institute of Mathematics and Centre for Medical and Biotechnical Research, National Academy of Sciences of Ukraine, Tereshchenkivska St. 3, 01030, Kyiv, Ukraine}

\begin{abstract}

We demonstrate that chimera behavior can be observed in small networks consisting of three identical oscillators, with mutual all-to-all coupling.
Three different types of chimeras, characterized by the coexistence of two coherent oscillators and one incoherent oscillator (i.e. rotating with another frequency) have been identified, where the oscillators show periodic (two types) and chaotic (one type) behaviors. Typical bifurcations at the transitions from full synchronization to chimera states and between different types of chimeras have been described. Parameter regions for the chimera states are obtained in the form of Arnold tongues, issued from a singular parameter point. Our analysis suggests that chimera states can be observed in small networks, relevant to various real-world systems.

\end{abstract}
\maketitle

Chimera states are spatiotemporal patterns consisting of spatially separated domains of coherent (synchronized) and incoherent (desynchronized) behavior, which appear in the networks of identical units. The original discovery in a network of phase oscillators [1-3] has sparked a tremendous activity of first theoretical studies \cite{amsw2008,l2009,mls2010,m2010,owm2010,omhs2011, maslm2014,epr2014,khskk2016} and next experimental observations \cite{hmrhos2012,tns2012,mtfh2013,sskg2014,lpm2013,kkwcm2014}.  In real-world systems, chimera states might play role in understanding of complex behavior in biological (modular neural networks \cite{network}, the unihemispheric sleep of birds and dolphins \cite{birds}, epileptic seizures \cite{epilepsy}), engineering (power grids \cite{grid,pshmr2014}) and social \cite{social} systems. More references can be found in two recent review papers \cite{pa2015,s2016}. 

Chimera states are typically observed in the large networks of different topologies, but recently it has been suggested that they can also be observed in small networks \cite{ab2015,paal2016,ba2016,b2016,bzsl2015}. Ashwin \& Burylko \cite{ab2015} have defined a {\it weak chimera state} as one referring to a trajectory in which two or more oscillators are frequency synchronized and one or more oscillators drift in phase and oscillate with different mean frequency with respect to the synchronized group.  First, it has been observed that these states can exist in small networks of as few as 4 phase oscillators \cite{ab2015,paal2016,ba2016,b2016} and also in the model of semiconductor lasers \cite{bzsl2015}.  Experimentally, chimera states of this type have been recently observed in small networks of optoelectronic oscillators \cite{Roy} as well as coupled pendula \cite{dgwpmk2016,wcmk2016}.

In the Letter, we show that the weak chimera patterns which are characterized by two frequency synchronized oscillators  and one evolving with different frequency can be observed in the networks of 3 identical nodes. As the proof of the concept we use a network of Kuramoto oscillators with inertia. We identify three different types of chimeras, namely (i) in-phase chimeras in which coherent oscillators are phase synchronized and incoherent one rotates with a different frequency, (ii) anti-phase chimeras in which coherent oscillators alternates with respect to each other and incoherent one oscillates with a different frequency, (iii) chaotic chimeras in which two oscillators are synchronized in frequency and the third one is not while the trajectories behavior is chaotic. 

We consider a ring of $N=3$ coupled pendulum-like nodes. Coupling is introduced in such a way that each
pendulum is connected to both its neighbors to the left and to the right with equal strength. The phase of each pendulum is described as follows:

\begin{equation}
m\ddot{\theta_{i}}+\varepsilon\dot{\theta_{i}}=\omega+\dfrac{\mu}{N}\sum_{j=1}^{N}\sin(\theta_{j}-\theta_{i}-\alpha),\label{eq:} 
\end{equation}
\noindent
where $i=1,...,N$,~ $\alpha$ is a phase lag,  $\mu$ is a coupling strength, $m$, $\varepsilon$, and $\omega$ are mass, damping and natural frequency of a single pendulum respectively. Eq.(\ref{eq:}) may be interpreted as an extension of Kuramoto model to the second-order differential equations, known as  Kuramoto model with inertia \cite{e1991,bkhb2014,onbt2014, jmk2015}). In our numerical simulations, we consider $m=1.0$, $\varepsilon=0.1$, and $\omega=0$ and explore the complex dynamics of the model (1) varying parameters $\mu>0$ and $0<\alpha<\pi$.  

Eq.(1) with $N=3$ is a 6-dimensional system of differential equations, but its effective dynamics is 4-dimensional given  in the following form:

\begin{align}
m\ddot{\eta_i}+\varepsilon\dot{\eta_i}=-\dfrac{\mu}{3} &  ( 2\cos\alpha\sin\eta_i+\\ \nonumber
+&\sin(\eta_{i+1}+\alpha)+\sin(\eta_i-\eta_{i+1}-\alpha)),
\end{align}
\noindent
where $\eta_1=\theta_1-\theta_2$, $\eta_2=\theta_1-\theta_3$ and $i=1,2$ . Equilibrium $(\eta_1,\eta_2)=(0,0)$ of the reduced system (2) corresponds to fully synchronized behavior in Eq.(1), where the phases of three pendula coincide and rotate with constant velocity $-(\mu/\varepsilon)\sin\alpha$.  For all $\mu>0$ the synchronized state is stable for $\alpha<\pi/2$ and unstable otherwise.  Four  eigenvalues of the equilibrium are the roots of characteristic equation $\left( \lambda^2+\dfrac{\varepsilon}{m}\lambda+\dfrac{\mu\cos\alpha}{m}\right) ^2=0$, and two remaining ones are respectively equal to $0$ and $-\mu/\varepsilon$. Therefore, chimera states reported below for $\alpha<\pi/2$ always co-exist in the phase space with fully synchronized state.  The second equilibrium of the reduced system (2) is equal to ($2\pi/3,-2\pi/3$), and corresponds to the stationary splay state $\theta_1=0,  \theta_2=2\pi/3,  \theta_3=-2\pi/3$ of Eq.(1). It is stable in a region around $\alpha=\pi$, loosing stability in a Hopf bifurcation as  $\alpha$ decreases.
  
Results of direct numerical simulation of system (1) are presented in Fig.1(a,b) in the two-parameter plane of the phase shift $\alpha$ and coupling strength $\mu$.  This figure reveals the appearance of regions of different chimera states, shown in shading (color), at intermediate values of $\alpha$ between $0$ and $\pi$.  Alternatively, if the pendula interact with  $\alpha$ close to $0$ or $\pi$, the collective behavior is synchronized in frequency, given by full synchronization or stationary splay state, respectively. 
Another peculiarity is that the region in Fig. 1(a) is detached from the $\mu=0$ level indicating that chimera states in Eq. (1) cannot exist for $\mu<0.015$.   Hence, the observed chimera states are not  a continuation of uncoupled multistability as in \cite{ab2015,paal2016,ba2016,b2016,bzsl2015,dgwpmk2016}. In contrary, the chimeras in system (1) appear due to non-vanishing pendula interaction and arise with increase of coupling strength $\mu$ in the homoclinic bifurcation (if $\alpha<\pi/2$) or in other bifurcations (if $\alpha>\pi/2$). With the further increase of $\mu$, the size of the chimera region grows, filling eventually the whole $\alpha$-interval between $0$ and $\pi$. 

\begin{figure}[ht]
	\begin{center}
		\epsfig{file=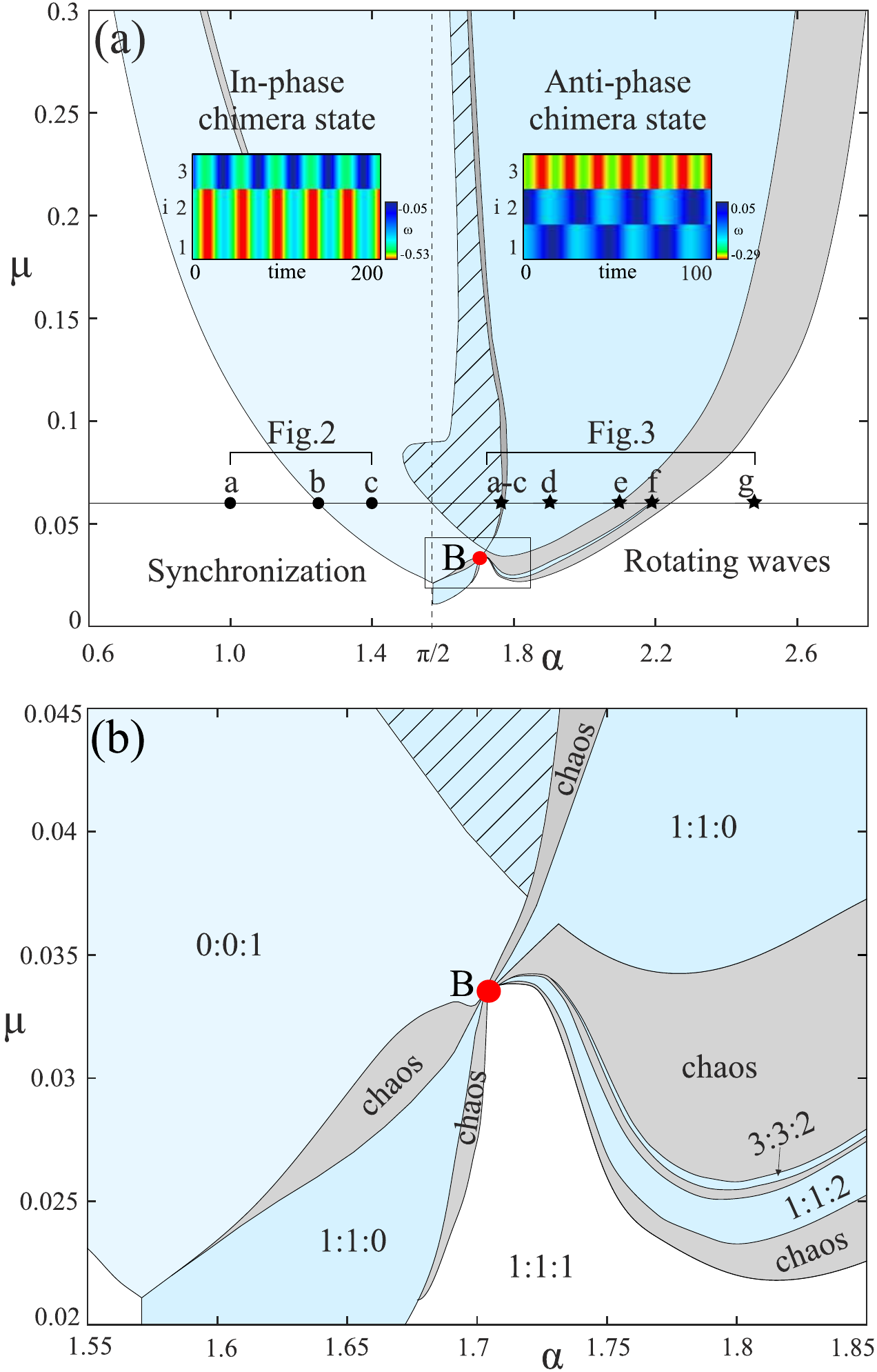,width=1.0\linewidth}
		\caption{(color online) (a) Dependence of possible spatio-temporal patterns on parameters
$\alpha$ and $\mu$ in system (1) . Frequency time plots of typical states are shown in the insets, (b) enlargement of the vicinity of the tongues origin.}
	\end{center}
\end{figure}

Parameter region in Fig.1(a) consists of two major domains, of {\it in-phase} chimera states (shown in light blue in the left side) and  {\it anti-phase} chimera states (darker blue region in the right side). These domains intersect along the dashed vertical strip in the middle (close to $\pi/2$).  By the in-phase chimera we mean the behaviour in Eq.(1), when two oscillators are fully synchronized ($\theta_1=\theta_2$),  but the third one $\theta_3$ is desynchronized and rotates with a different average frequency. The anti-phase chimera means a similar frequency behaviour however phases $\theta_1$ and $\theta_2$ are not equal, but alternate with respect to each other with phase difference around $\pi$. Typical frequency-time plots of the in-phase and anti-phase chimera states are shown in insets in Fig.1(a). The third characteristic  behaviour is a {\it chaotic} chimera state found in tiny parameter strips at the transitions between different states. All oscillators behave chaotically in time preserving, nevertheless, the frequency synchronization of two of them and desynchronization of the third one.   

Bifurcation diagram in Fig.1(a) has regions of synchronization similar to Arnold tongues of a three-dimensional torus $T^3$ for phase variables $\theta_1, \theta_2,$ $\theta_3$.  Our simulations show that tongues of synchronization originate from a singular parameter point $B$ equal   $\alpha=1.70111$ and  $\mu=0.03317$.   
Detailed bifurcation structure in the neighborhood of the singular point  $B$ is shown on the enlargement in Fig. 1(b). The largest tongue is for the in-phase chimera state (see left snapshot in Fig.1(a)) originates from $B$ in the left-up direction; we denote it by $(0:0:1)$ indicating, that one oscillator (let it be the third $\theta_3$) rotates faster then the two others synchronized.  The anti-phase chimera state (see right snapshot in Fig.1(a)) exists inside two tongues which originate from $B$ in the opposite directions: one goes right-up and the other (smaller) left-down. These tongues are denoted in Fig. 1(b) by $(1:1:0)$ to point out that in this case two synchronized oscillators rotates faster then the third one desynchronized. Additionally, two narrow tongues of the other modality are shown to the right, denoted by $(1:1:2)$ and $(3:3:2)$, which indicate relative numbers of phase slips of the oscillators.  Further increase of the calculations precision yields additional thin high-order resonant tongues (not shown in Fig. 1(b)). 

Each region of synchronization is characterized by the Poincare winding numbers, which determine the average frequencies of the individual oscillators $(\bar{\omega_1}, \bar{\omega_2}, \bar{\omega_3})$.   Chimera states appear in the tongues where two of the rotating numbers coincide but the third one is different, i.e., $\bar{\omega_1}=\bar{\omega_2}\neq\bar{\omega_3}$ (see \cite{ab2015} for definition).  Four tongues mentioned above fulfill this property.  On the other hand, chimeras do not arise if all oscillators are  frequency synchronized, i.e. $\bar{\omega_1}=\bar{\omega_2}=\bar{\omega_3}$, or rotate with three different average frequencies, i.e. $\bar{\omega_1}\neq\bar{\omega_2}\neq\bar{\omega_3}$ . An example of the non-chimera behavior is given as rotating waves in the tongue $(1:1:1)$ (blank region in Fig.1(a,b) for $\alpha>\pi/2$), as well as many other tiny tongues inside the regions of chaoticity, which also originate from the singular parameter point $B$.  Indeed, each chaotic region indicated in Fig.1(a,b) is filled by many thin regions of synchronization (windows of periodicity), while each of them is characterized by the winding numbers $(\bar{\omega_1}, \bar{\omega_2}, \bar{\omega_3})$. For chimera state, we need that two of the numbers coincide but the third one does not. Each such region of synchronization contains normally a chaotic part arising in a sequence of bifurcations (like in windows for 1-dimensional logistic map). Then, this chaotic behaviour inside the window of periodicity corresponds to chaotic chimera state when the window winding numbers fulfill the chimera condition  $\bar{\omega_1}=\bar{\omega_2}\neq\bar{\omega_3}$.  To our surprise, chaotic chimera behaviour in model (1) appears to be quite common, not as rare as has been expected. 

A  typical scenario for the appearance/disappearance of the chimera states in model (1) is illustrated in Fig.2(a-c).  We fix coupling strength $\mu=0.06$ and increase the phase lag parameter $\alpha$ along the horizontal line with bold points in Fig.1(a). First, if $\alpha$ is small enough, phases of all three oscillators coincide $\theta_{1}=\theta_{2}=\theta_{3}$ and rotate with constant velocity $\dot{\theta_{i}}=-(\mu/\varepsilon)\sin\alpha$, as shown in the phase-time plot in the right panel of Fig.2(a). The increase of $\alpha$ causes a homoclinic bifurcation, which occurs at some $\alpha=\alpha_0 \approx 1.247697$ producing an in-phase chimera state with the property $\theta_{1}=\theta_{2}\neq\theta_{3}$ and $\bar{\omega_1}=\bar{\omega_2}\neq\bar{\omega_3}$ as shown in the phase-time plot in the right panel of Fig. 2(c). The system dynamics becomes two-dimensional given by an equation in the manifold $\theta_1=\theta_2$: 
\begin{equation}
m\ddot{\eta}+\varepsilon\dot{\eta}=-\dfrac{\mu}{3}\left(2\cos\alpha\sin\eta+\sin(\eta-\alpha)+\sin\alpha\right),\label{eq:in}
\end{equation}
where $\eta=\theta_1-\theta_3$.  The phase portraits  in the  vicinity of homoclinic bifurcation parameter point $\alpha_0$ are shown in Fig.2(a-c).
\begin{figure}[ht]
	\begin{center}
		\epsfig{file=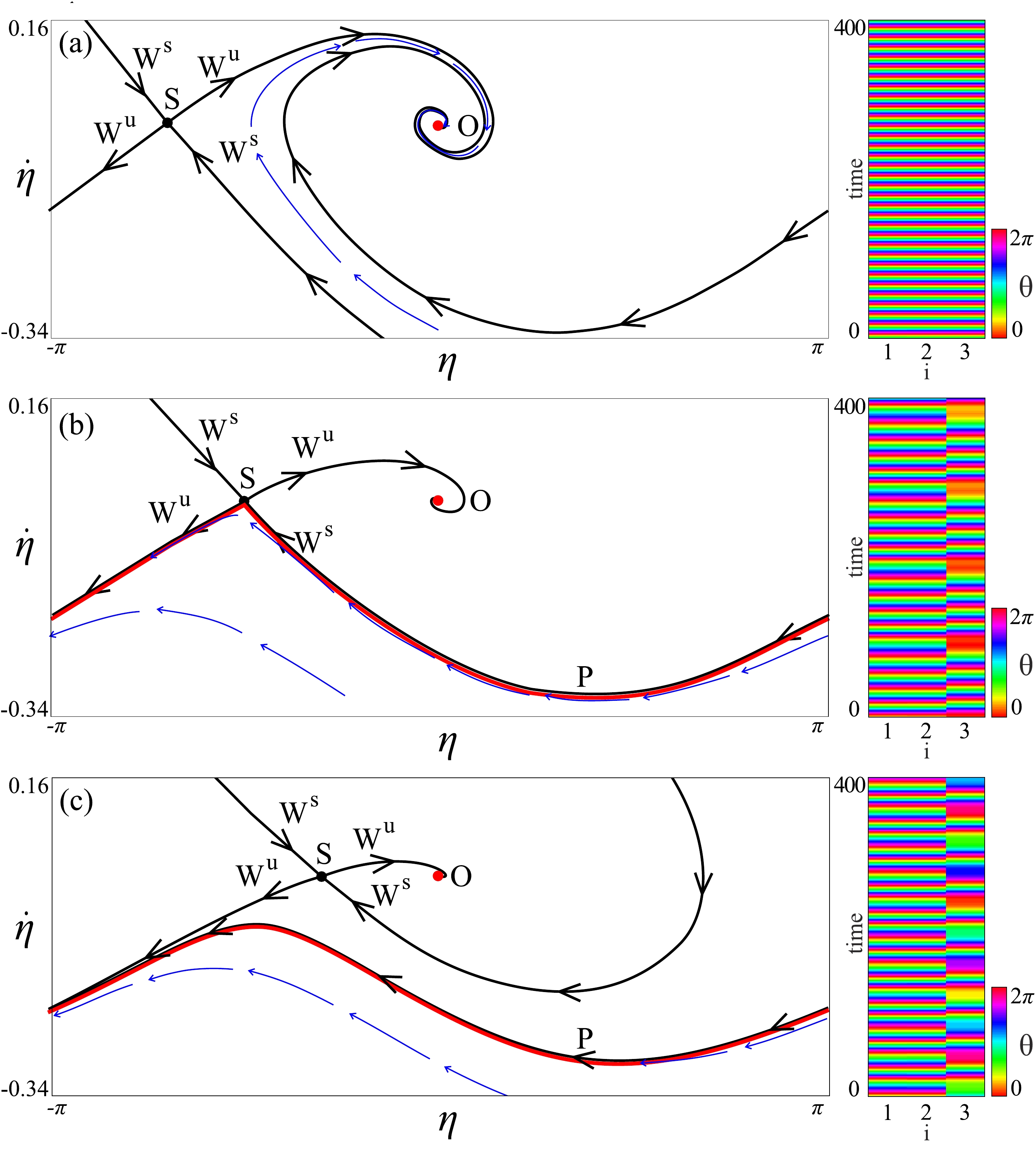,width=1.0\linewidth}
		\caption{(Color online). Phase portraits (left panels) and phase-time plots (rigth panels) in the vicinity of homoclinic bifurcation point in system (1): $\mu=0.06$; (a) before bifurcation $\alpha=1.0$, (b) bifurcation point  $\alpha=1.247697$, (c) after bifurcation $\alpha=1.4$. $\eta=\theta_1-\theta_3$. Equilibria are given by $O=(0,0) $ and $S= (\mathrm{atan2}(-6\tan(\alpha), -9+\tan(\alpha)^2),0)$, while $P$ stands for periodic orbit.}
	\end{center}
\end{figure}
For $\alpha<\alpha_0$,  all trajectories of Eq.(3) end up in the stable equilibrium $O=(0;0)$, except for the saddle point $S$ and two branches of its stable manifold $W^{s}$ [Fig.2(a)]. At $\alpha=\alpha_0$, the left branch of the unstable manifold $W^{u}$ hits the right branch of the stable manifold $W^{s}$ [Fig.2(b)]. A homoclinic orbit is created, where the behaviour is asymptotic to saddle $S$ in both directions $t\rightarrow\pm\infty$.  After the bifurcation, for $\alpha>\alpha_0$, a stable limit cycle $P$ is born detaching from the homoclinic orbit  [Fig.2(c)]. The behavior along the limit cycle $P$ corresponds to the in-phase chimera state. Indeed, when rotating along the cycle the phase difference $\eta=\theta_1-\theta_3$  between the first and the third oscillators grows while the difference $\theta_1-\theta_2$ remains bounded (equals zero) and hence the third oscillators rotates with different frequency.  Note, that chimera state born in the homoclinic bifurcation co-exists with the fully synchronized rotation given by stable equilibria $O$ of Eq.(3). When $\alpha$ only slightly exceeds $\alpha_0$, basin of attraction of the chimera state is small and hard to catch from random initial conditions.   With increase of $\alpha$, the chimera basin swells and fills up eventually almost the whole space volume as $\alpha\rightarrow\pi/2$. 

\begin{figure}[ht]
	\begin{center}
		\epsfig{file=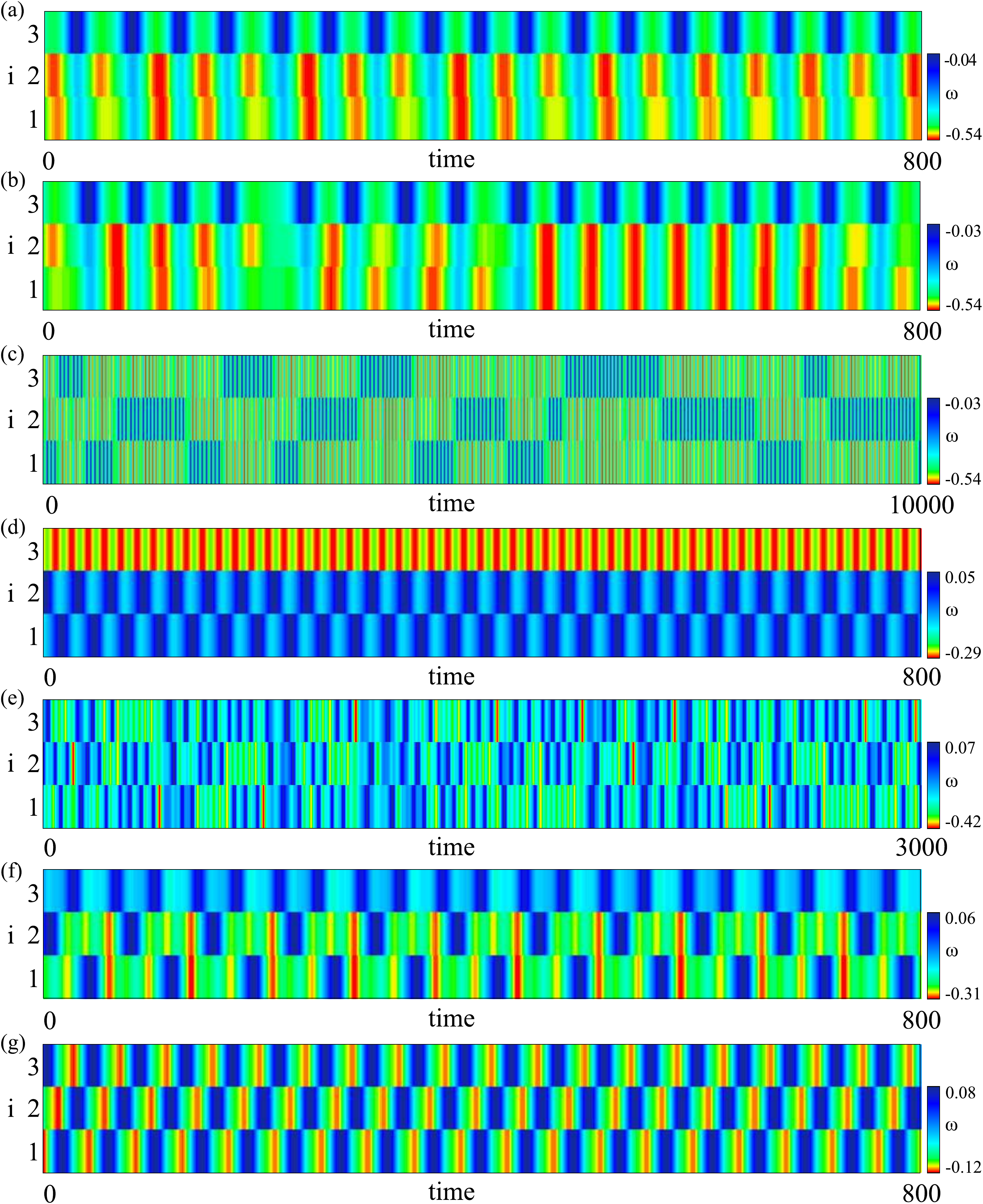,width=1.0\linewidth}
		\caption{(Color online). Frequency time plots of system (1): $\mu$ = 0.06, (a) $\alpha=1.760$ - imperfect chimera; (b) $\alpha=1.765$ - chaotic chimera; (c) $\alpha=1.7665$ - heteroclinic cycling; (d) $\alpha=1.90$ - anti-phase chimera;
		(e) $\alpha=2.10$ - chaotic chimera; (f) $\alpha=2.182$ - (1;1;2) anti-phase chimera; (g) $\alpha=2.47$ - rotating waves. }
	\end{center}
\end{figure}

\begin{figure}[ht]
	\begin{center}
		\epsfig{file=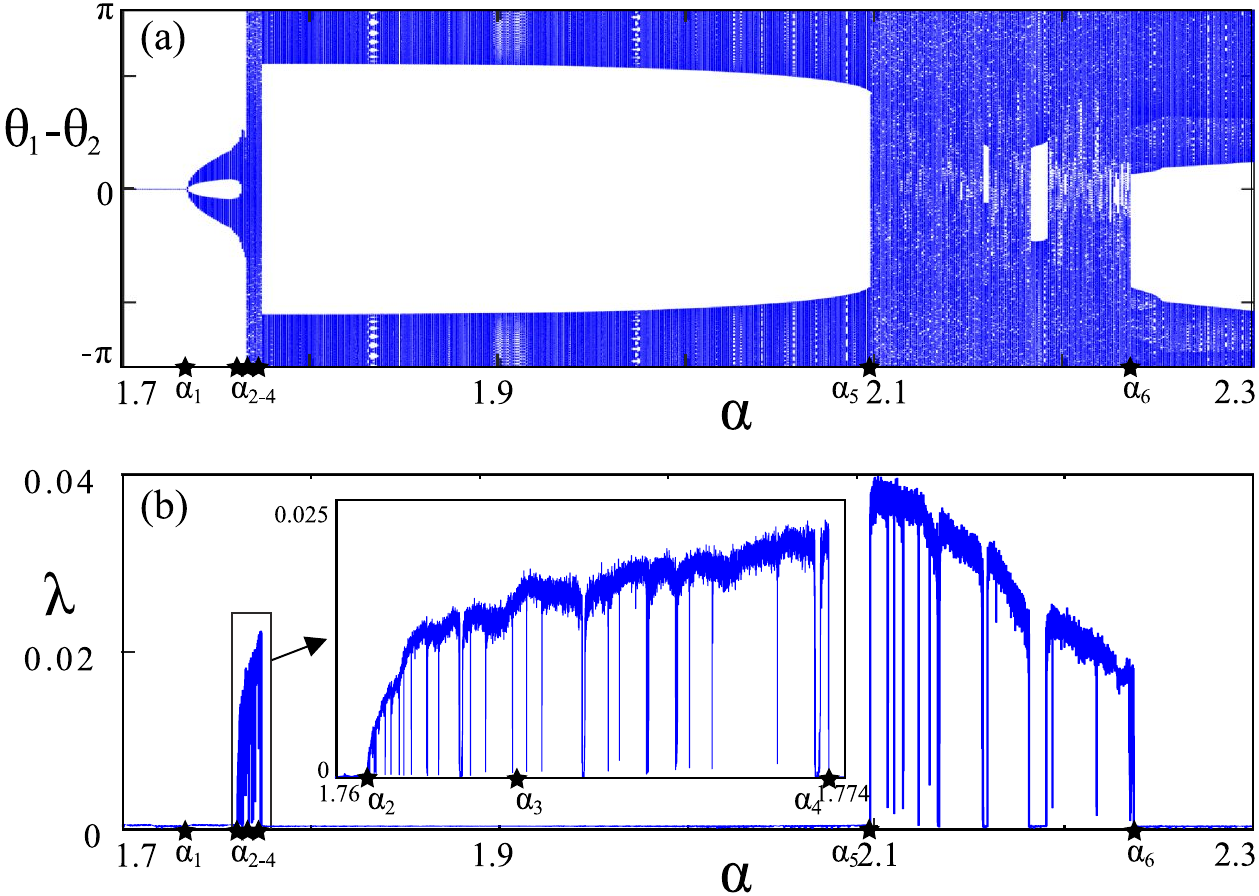,width=1.0\linewidth}
		\caption{(color online) (a) Bifurcation diagram and (b) maximal Lyapunov exponent for system (1): $\mu$ = 0.06.}
	\end{center}
\end{figure}

Further changes of the system dynamics are caused by the appearance of anti-phase chimera state at the left boundary of the multistability region, shown dashed in Fig.1(a). It is born in a saddle-node bifurcation and exists thereafter at a large $\alpha$-interval,  up to the right hand side chaos region in Fig.1(a).  We will return to this state later, but now, let us follow the in-phase state for $\alpha$ beyond $\pi/2$, as illustrated in Fig. 3(a-g) and 4(a,b).  First, the in-phase chimera state loses its 'perfection' in the sense that phases of the synchronized oscillators $\theta_1$ and $\theta_2$ become not equal, however they still rotate with the same average frequency.  This happens in a pitchfork bifurcation at $\alpha_1$ as shown in Fig. 4(a).  With the further increase of $\alpha$, the behavior of the phase difference $\theta_1-\theta_2$ is tangling and becomes chaotic at $\alpha_2$,  preserving nevertheless the weak chimera property $\bar{\omega_1}=\bar{\omega_2}\neq\bar{\omega_3}$. Chaotic chimera state shown in Fig.3(b) is born, and  exists in $\alpha$-interval $(\alpha_2,\alpha_3)$ as confirmed by  the plot of maximum Lyapunov exponent shown in Fig.4(b). At $\alpha=\alpha_3$, chaotic chimera loses its stability transforming into a chaotic saddle. Thus, chimera state disappears and the system behavior develops then in the form of {\it heteroclinic cycling} between the three symmetric chaotic saddles born at $\alpha_3$ (existing due to the permutation symmetry of the model),  as shown in Fig.3(c).
System trajectory  is 'jumping'  between the three chimera-saddles and generally, chimera states do not exists in this case. However, it may happen that the trajectory spends equal time near two of the chimera-saddles and less (or more) time near the third one, satisfying hereby the chimera criterion $\bar{\omega_1}=\bar{\omega_2}\neq\bar{\omega_3}$. This kind of chimera-like heteroclinic cycling  can be found in narrow windows of periodicity inside the chaotic regions, we leave it for future study.

The heteroclinic cycling between the three chimera-saddles lasts up to some $\alpha=\alpha_4$, where it disappears in a crisis bifurcation. The system behaviour drops on the anti-phase chimera state [Fig.3(d)] born earlier at the left border of the multistability region close to $\pi/2$.  This anti-phase chimera state persists with further increase of $\alpha$ up to $\alpha_5$, where it disappears in an inverse saddle-node. Parameter point $\alpha,\mu$ enters the second, more pronounced
region of chaoticity and the chimera states cease to exist. After, system behavior develops in the form of fast disordered jumping between different saddle-type rotations born originally in the singular  parameter point $B$, as shown in Fig. 3(e). On the other hand, when entering deeper into the chaotic region with more increase of $\alpha$, numerous windows of periodicity become more apparent. Most of the windows are invisibly thin, only a few can be detected by standard simulations. An example is given by the (1:1:2)-tongue (shown in Fig.1(a,b)) which intersects the $\mu=0.06$ level through the $\alpha$-interval approximately equal $(2.1743, 2.1812)$ . The behaviour here is shown in Fig. 3(f).  It represents an anti-phase chimera state, however  different from those in the wide (1:1:0)-tongue between the two chaotic regions, as the non-synchronized oscillator rotates faster in this case than two others synchronized.  Similarly, chimeras can arise in other, microscopic windows as soon as two winding numbers coincide but the third one does not. 
As $\alpha$ increases further, chaotic region ends at some $\alpha=\alpha_6$, and the system dynamics dive into nonlinear rotating waves [Fig. 3(g)] leading eventually to the stationary splay state as $\alpha$ further increases to $\pi$.

In conclusions, we showed that chimera states typically arise in the Kuramoto model with inertia,  at the transition from coherence to rotating waves. We described the transition in details for the coupling strength $\mu$ above the singular point $B$, fixing $\mu=0.06$ and varying $\alpha$ from $0$ till $\pi$.  Our simulations have confirmed, that similar scenario takes place for other $\mu$, also below $B$. The sequence of bifurcations can be different but the main features remain the same:  
(i) {\it in-phase chimera state} appears in a homoclinic bifurcation with an increase of phase shift $\alpha$ (or coupling strength $\mu$) ;  {\it anti-phase chimera state} arises at further variation of $\alpha$ in a saddle-node bifurcation;  
(ii)  secondary chimera states arise inside the chaos regions when parameter point enter the windows of periodicity (Arnold tongues) with the winding numbers satisfying the Ashwin-Burylko criterion $\bar{\omega_1}=\bar{\omega_2}\neq\bar{\omega_3}$;  the behaviour inside the tongue first is regular (periodic or quasiperiodic), then becomes chaotic in a sequence of bifurcations producing {\it chaotic chimera state};  
(iii) when exiting the tongue of synchronization, chimera state transforms into a saddle-chimera, giving rise to heteroclinic cycling characterizing the next level of the chimera complexity.  
We suggest, that this  indicate a common, probably universal scenario for the chimera state appearance in oscillatory networks of different nature, due to presence of inertia.

\textbf{Acknowledgement:} This work has been supported by the Polish National Science Centre, MAESTRO Programme - Project No 2013/08/A/ST8/00/780.

\end{document}